\documentclass[conference]{IEEEtran}
\IEEEoverridecommandlockouts
\usepackage{cite}
\usepackage{amsmath,amssymb,amsfonts}
\usepackage{algorithmic}
\usepackage{graphicx}
\usepackage{textcomp}
\usepackage{physics}
\usepackage{xcolor}
\def\BibTeX{{\rm B\kern-.05em{\sc i\kern-.025em b}\kern-.08em
    T\kern-.1667em\lower.7ex\hbox{E}\kern-.125emX}}
\begin{document}

\title{Quantum Machine Learning: Performance and Security Implications in Real-World Applications}

\author{\IEEEauthorblockN{Zhengping Jay Luo}
\IEEEauthorblockA{Department of Computer Science and Physics\\Rider University\\
Lawrenceville, New Jersey, 08648\\
Email: zluo@rider.edu}
\and
\IEEEauthorblockN{Tyler Stewart}
\IEEEauthorblockA{Department of Computer Science and Physics\\Rider University\\
Lawrenceville, New Jersey, 08648\\
Email: stewartty@rider.edu}
\and
\IEEEauthorblockN{Mourya Narasareddygari}
\IEEEauthorblockA{Department of Computer Science and Physics\\Rider University\\
Lawrenceville, New Jersey, 08648\\
Email: mnarasaredd@rider.edu}
\and
\IEEEauthorblockN{Rui Duan}
\IEEEauthorblockA{School of Science and Engineering\\University of Missouri-Kansas City\\
Kansas City, Missouri, 64110\\
Email: rdhk9@umkc.edu}
\and
\IEEEauthorblockN{Shangqing Zhao}
\IEEEauthorblockA{School of Computer Science\\University of Oklahoma\\
Tulsa, Oklahoma, 74135\\
Email: shangqing@ou.edu}
}

\maketitle

\begin{abstract}
Quantum computing has garnered significant attention in recent years from both academia and industry due to its potential to achieve a "quantum advantage" over classical computers. The advent of quantum computing introduces new challenges for security and privacy. This poster explores the performance and security implications of quantum computing through a case study of machine learning in a real-world application. We compare the performance of quantum machine learning (QML) algorithms to their classical counterparts using the Alzheimer’s disease dataset. Our results indicate that QML algorithms show promising potential while they still have not surpassed classical algorithms in terms of learning capability and convergence difficulty, and running quantum algorithms through simulations on classical computers requires significantly large memory space and CPU time. Our study also indicates that QMLs have inherited vulnerabilities from classical machine learning algorithms while also introduce new attack vectors.  
 \end{abstract}

\begin{IEEEkeywords}
Quantum Security, Quantum Machine Learning, Quantum Advantage, Attack Vectors
\end{IEEEkeywords}

\section{Introduction}

Since Richard Feynman first proposed the idea of harnessing quantum physics to build quantum computers more than 40 years ago \cite{preskill2023quantum}, significant breakthroughs and progress have steadily been made toward realizing Feynman's vision and achieving "quantum advantage."  With the development of quantum computing technologies, a natural question to ask is how to protect the security and privacy in a quantum age? We conduct a case study regarding performance and security implications of machine learning (ML) algorithms in quantum computing. In recent years, numerous quantum machine learning (QML) proposals have been published \cite{cong2019quantum,cerezo2021variational,havlivcek2019supervised}, paving the way for unleashing the full potential of ML algorithms on quantum computers.

In this poster, we want to know the performance of QML algorithms compared to their classical counterparts on a real-world dataset, and the corresponding security implications. We conduct a comparative study of the performance of two major types of classical machine learning (CML) algorithms—support vector machines (SVMs) and multi-layer perceptron (MLP) classifiers—and their corresponding quantum versions, including quantum support vector machines (QSVMs), variational quantum algorithms (VQAs) and quantum convolutional neural networks (QCNNs). Our comparison and analysis are based on the real-world Alzheimer's disease dataset \cite{rabie_el_kharoua_2024}. Then we'll discuss the potential security implications of the QML algorithms, including the inherited vulnerabilities from their classical counterparts and the new introduced attack vectors.

%Traditional cryptographic systems, which form the backbone of secure communication and data protection, rely on the computational difficulty of certain mathematical problems, such as integer factorization and discrete logarithms. These problems are effectively insurmountable for classical computers. However, quantum computers can solve many of the computationally difficult problems on classical computers exponentially faster \cite{shor1999polynomial, grover1997quantum}, rendering many of today's cryptographic protocols vulnerable to decryption.

\section{Quantum Machine Learning}
QML is the quantum counterpart to CML. In CML, there are two main categories: supervised and unsupervised learning. This poster focuses on supervised learning. Kernel methods, such as SVMs, and neural network-based methods, including MLPs and Convolutional Neural Networks (CNNs), are among the most renowned families of supervised learning algorithms that have achieved significant success on classical computers.

The idea of kernel methods is to map the classification problem non-linearly to a high-dimensional feature space, making data easier to separate.  QSVMs \cite{havlivcek2019supervised} use the quantum state space as feature space by mapping classically processed data non-linearly to a quantum state $\Phi$, i.e., $\overrightarrow{x} \in \Omega \rightarrow \ket{\Phi(\overrightarrow{x})} \bra{\Phi(\overrightarrow{x})}$, in which $\Omega \subset \mathbb{R}^{d}$. This allows leveraging quantum feature maps to perform the kernel trick, with the feature vector kernel represented as $\mathit{K}(\overrightarrow{x},\overrightarrow{z})= |\bra{\Phi(\overrightarrow{x})}\ket{\Phi(\overrightarrow{z})}|^2$, where $\Phi(\overrightarrow{x})$ and $\Phi(\overrightarrow{z})$ are quantum feature maps. According to Havl{\'\i}{\v{c}}ek et al.\cite{havlivcek2019supervised}, a quantum advantage for QSVMs can only be achieved for feature maps with a kernel that is hard to estimate classically.

VQAs, the quantum analog of MLP neural networks, are designed to run on quantum computers using a classical optimizer to train parameterized quantum circuits. They are currently considered the best hope and candidate for achieving quantum advantage on noisy intermediate-scale quantum (NISQ) devices\cite{cerezo2021variational}. The principle behind VQAs is to encode the problem into a cost function $\mathit{C}$ with a set of parameters $\boldsymbol \theta$. The optimizer's task is to solve $\boldsymbol{\theta}^{\ast} = \arg \underset{\boldsymbol{\theta}}{\min}\mathit{C}(\boldsymbol{\theta})$,where $\mathit{C}(\boldsymbol{\theta})=f(\{\rho_k\},\{\mathit{O}_k\},\mathit{U}(\boldsymbol{\theta}))$. Here $\mathit{U}(\boldsymbol{\theta})$ denotes a unitary operation, $\{\rho_k\}$ are the input states and $\{\mathit{O}_k\}$ denotes a set of observables. When the optimization task converges, the final ansatz and the corresponding learned model can be used for future applications. %However, there are multiple concerns about the practical usage of VQAs, such as the barren plateau phenomenon in the cost function landscape and the efficiency of estimating the expectation values of the cost function \cite{cerezo2021variational}.

Another notable proposal for QML models is the QCNN developed by Cong et al. \cite{cong2019quantum}. QCNNs extend the key properties of classical CNNs to the quantum field by incorporating all major components of CNNs, including convolutional layers and pooling layers, with the input transformed into quantum states. Convolution, pooling operations, and fully connected layers are implemented using unitary rotations. The nonlinearity in QCNNs arises from reducing the degrees of freedom. The variational parameters required in QCNNs are only $\mathit{O}(\text{log}(\mathcal{N}))$, given an input size of $\mathcal{N}$ bits, which allows for efficient training and implementation.

From the models introduced above, we know that there are many structural similarities between QML models and their corresponding CML models. Thus from a security perspective, QML models will inevitably inherit many of the vulnerabilities in CML, such as the crafted adversarial examples attack. However, as stated in \cite{franco2024predominant}, it will be more difficult to perform robust adversarial training due to the increased dimensions of the space used in QML, making it more sensitive to minor perturbations near the decision boundary. 

\section{Experimental Results and Analysis}
We compared the performance of the aforementioned representative QML algorithms to their classical counterparts on the real-world Alzheimer's disease dataset\cite{rabie_el_kharoua_2024}, which is a public dataset containing health information for 2,149 patients, including demographic details, lifestyle factors, medical history, and more. We used this dataset to train learning models to predict whether a patient will be diagnosed with Alzheimer's disease, framing it as a binary classification problem. We implemented our learning models using the scikit-learn library \cite{scikit-learn} and the IBM Qiskit software stack.\cite{qiskit2024}. we employed two CML models—SVM and MLP—to train models. Additionally, we trained three types of QML models for performance comparison: QSVM, VQA, and QCNN, respectively.

We first studied the amount of training data required to achieve a "reasonably" good learning model using SVM and MLP models. Our findings indicate that as the size of the training data increases, the performance, measured by prediction accuracy on the test dataset, improves moderately from 80\% (when using 10\% of the dataset for training) to 87\% (when using 90\% for training). Notably, among the 32 features in the dataset, the most influential factor contributing to the diagnosis of Alzheimer's disease is "Memory Complaints," providing an intuitive way to assess whether a patient has Alzheimer's with high probability. The best accuracy we achieved using either SVM or MLP was 87\% on the testing data, with 90\% of the data allocated for training.

When trained using QML algorithms, our models generally did not surpass their classical counterparts in both time and accuracy, except for the QCNN model, which has the same prediction accuracy as the SVM model. This is understandable, as CML algorithms have been refined and optimized over many years by numerous researchers. Accessing real-world quantum computers remains challenging; therefore, we ran our algorithms on local computers with simulations, which typically result in significantly longer training times compared to classical methods. For example, training a classical SVM model takes only 0.03 seconds of CPU time, while training the same-sized QSVM model on a local computer via simulation takes 132.07 seconds—over 4,000 times longer in terms of time cost. We anticipate that as real-world quantum computers become more accessible, this gap will be narrowed and potentially reversed.

Given above experimental results, we can arguably infer that it will be easier for adversarial example attacks to succeed when the prediction from QML models are less accurate than CML models, often resulted by defective decision boundaries. Further more, in quantum scenarios, new attack vectors could also be introduced, including exploiting quantum noise to degrade the performance of quantum models \cite{franco2024predominant}.

\section{Conclusion}
Security concerns of quantum computing has been an emerging problem. We employed ML as a case study to explore the performance and security implications of quantum computing. Our results indicate that QML algorithms such as QSVM, VQA, QCNN models still have not surpassed CML models in terms of learning performance on our real-world dataset in a simulated environment. The security implications of quantum computing are multi-faceted. In our future work, we'll continue to explore more security implications of quantum computing in ML domain. 

\bibliographystyle{IEEEtran}
\bibliography{IEEEabrv}

\end{document}